\documentclass{article}

\usepackage{lineno,hyperref}
\usepackage{graphicx}
\usepackage{amsmath}
\usepackage{amssymb}
\usepackage{multirow}
\usepackage{color,soul}

\begin{document}

\title{Non-relativistic Arbitrary \textit{l}-states of Quarkonium through Asymptotic Iteration Method}


\author{Hakan Ciftci$^{1}$\thanks{Corresponding author. E-mail:hciftci@gazi.edu.tr}, \ Hasan Fatih Kisoglu$^{2}$\thanks{Corresponding author. E-mail:hasanfatihk@mersin.edu.tr}\\
	$^{1}${Gazi \"{U}niversitesi, Fen-Edebiyat Fakültesi, Fizik B\"{o}l\"{u}m\"{u}, 06500 Teknikokullar-Ankara, Turkey}\\  
	$^{2}${Department of Basic Sciences, Faculty of Maritime, Mersin University, Mersin, Turkey}}


\date{\today}
\maketitle

\begin{abstract}
The energy eigenvalues with any $\textit{l}\neq 0$ states and mass of heavy quark-antiquark system (quarkonium) are obtained by using Asymptotic Iteration Method in the view of non-relativistic quantum chromodynamics, in which the quarks are considered as spinless for easiness, and are bounded by Cornell potential. A semi-analytical formula for energy eigenvalues and mass is achieved via the method in scope of the perturbation theory. The accuracy of this formula is checked by comparing the eigenvalues with the ones numerically obtained in this study, and with exact ones in literature. Furthermore, semi-analytical formula is applied to c\={c}, b\={b} and c\={b} meson systems for comparing the masses with the experimental data.
\end{abstract}



\section{Introduction}

Investigation of an atomic or sub-atomic system is done by achieving an energy spectrum of the system. This is carried out for the events in which the system is bounded by a potential function. The eigenvalues (or eigenenergies) of Hamiltonian of this system is obtained for a given potential function. In order to do this, various mathematical methods are used in quantum mechanics. One of these, named Asymptotic Iteration Method (AIM), has been commonly used since 2003 \cite{orj}. AIM can be used for analytically, as well as numerically (or approximately) solvable problems \cite{trigonometric,construction,eolgar,makale,aimdirac,fernandez}. Moreover, it can be used for obtaining the perturbative energy eigenvalues of the system without any need of the unperturbative eigenstate \cite{pert,makale2}.

As a sub-atomic system, a quarkonium that is composed of a heavy quark-antiquark (q\={q}) pair has attracted attention of particle physicists since the first half of 1970, and Refs. \cite{eichten1,eichten2,eichten3,chung,hamzavi} are just a few studies of them. In most of these studies, for easiness, the system is examined via Schr\"{o}dinger equation in non-relativistic quantum chromodynamics (NRQCD), assuming that the quarks are spinless \cite{alfredo,jacobs,grinstein,lucha}. Cornell potential is one of the potential functions that represent interactions between the quarks in such a q\={q} system. It is used for obtaining the mass and energy spectrum of the quarkonium and obtaining the hadron decay widths \cite{eichten1,eichten2,eichten3,evans}. Cornell potential is given as

\begin{equation}
V(r)=-\frac{A}{r}+B^{2}r  \label{intpot}
\end{equation}
where $\mathit{A}$ and $\mathit{B}$ are positive constants. As it is seen in Eq.(\ref{intpot}), Cornell potential has two parts: one is the Coulombic term and the other is the linear part. For obtaining the energy levels and mass of the quarkonium, $A$ and $B$, may be fitted to the first-few states. Therefore, the full spectrum of the quarkonium can be constructed through these potential parameters.

In literature, it is possible to find many studies in which the solutions of Schr\"{o}dinger equation for Cornell potential have been obtained. For example, in \cite{rlhall1984}, Hall has found an approximate energy formula to construct an energy spectrum of Schr\"{o}dinger equation for Cornell  potential, under some conditions. Jacobs et al. \cite{jacobs} have compared the eigenvalues of Schr\"{o}dinger and spinless Salpeter equations in the cases of Cornell potential and Wisconsin potential \cite{hagiwara}. Vega and friends have obtained, for \textit{l}=0 states, the energy spectrum, mass and wavefunctions at the origin for c\={c}, b\={b} and b\={c} mesons by using the usual variation method in the scope of super-symmetric quantum mechanics (SUSYQM) \cite{gozzi,cooper}, in Ref.\cite{alfredo}.  They have also compared their results with the exact ones in literature, and with the experimental data. 


In this study, we attempted to get the energy eigenvalues (for any $\textit{l}\neq 0$ states) and masses of heavy mesons by using Asymptotic Iteration Method in the view of NRQCD, in which the quarks are considered as spinless for easiness, and are bounded by Cornell potential. We achieved a semi-analytical formula for constructing the energy spectrum and obtaining the masses of the mesons, using the method in scope of the perturbation theory. The accuracy of this formula was cross-checked by comparing the eigenvalues with the ones numerically obtained in this study, and with the exact ones in literature. Furthermore, semi-analytical formula was applied to c\={c}, b\={b} and c\={b} heavy mesons for comparing the masses with the experimental data. 

AIM has been firstly applied to Schr\"{o}dinger equation for Cornell potential by Hall and Saad in Ref.\cite{hallnsaad}. They have used Airy function as an asymptotic form of the
wavefunction, and have got highly-accurate numerical results in their study. Alternatively, we obtained a semi-analytical mass-energy formula for quarkonium by having differential equation which gives polynomial solutions for asymptotic forms of the wavefunction of the system.

This paper is organized as follows: we give a short summary of AIM in Section 2, while Section 3 includes the main problem. In Section 4, we give numerical results for the eigenenergies, and obtain semi-analytical energy formula by applying perturbation theory to our problem in the view of AIM. Furthermore, in Section 4, we compare our energy spectrum and masses with the exact ones in literature, and with the experimental data.  Finally, Section 5 includes some comments about our results.

\section{The Asymptotic Iteration Method (AIM)}\label{sec:1}
According to tge organization of the paper, we summed up AIM in this section, while it is comprehensively introduced in Ref.\cite{orj}. The AIM is used to
solve second-order homogeneous linear differential equations in the following form

\begin{equation}
y^{\prime \prime }(x)=\lambda _{0}(x)y^{\prime }(x)+s_{0}(x)y(x)
\label{aimform}
\end{equation}
where $\lambda _{0}(x)$ and $s_{0}(x)$ have continuous derivatives in the
defined interval of the $\mathit{x}$ independent variable. If there is an asymtotic condition such as

\begin{equation}
\frac{s_{n}}{s_{n-1}}=\frac{\lambda _{n}}{\lambda _{n-1}}\equiv \alpha
\label{asp}
\end{equation}
for $n\in \mathbb{Z}^{+}$, where $n$ is large enough, the general solution of Eq.\ref{aimform}
is obtained as

\begin{equation}
y(x)=\exp \left( -\int\limits^{x}\alpha (t)dt\right)\left[
C_{2}+C_{1}\int\limits^{x}\exp \left( \int\limits^{t}\left( \lambda
_{0}(\tau)+2\alpha (\tau)\right) d\tau \right) dt\right] \label{aimsol}
\end{equation}
with the functions

\begin{eqnarray}
\lambda _{n} &=&\lambda _{n-1}^{\prime }+s_{n-1}+\lambda _{0}\lambda _{n-1}
\label{lames} \\
s_{n} &=&s_{n-1}^{\prime }+s_{0}\lambda _{n-1}  \notag
\end{eqnarray}

As a field of application, AIM can be used to deal with Schr\"{o}dinger equation (or energy eigenvalue problem) in mathematical physics. The eigenvalues can be obtained through the following quantization condition;

\begin{equation}
\delta _{n}(x, E)=s_{n}(x, E)\lambda _{n-1}(x,E)-\lambda _{n}(x,E)s_{n-1}(x,E)=0
\label{delta}
\end{equation}

If the energy eigenvaules ($E$) can be obtained from Eq.(\ref{delta}), independently from the $x$ variable, the problem is exactly solvable. In this case, the eigenvalue and eigenfunction of $n$th energy level can be derived in explicit algebraic form via $n$ iterations. However, there are limited numbers of suitable potentials for this case.

As for the approximately (or numerically) solvable problems, $\delta _{n}$ depends on both $\mathit{x}$ and $\mathit{E}$. In this case, an appropriate value, $\mathit{x}\equiv\mathit{x}_{0}$, should be determined to solve $\delta _{n}(x,E)=0$ with respect to $E$ \cite{trigonometric, makale2}. The energy eigenvalue of an $n$th level is obtained through $q$ iterations where $q\geq n$.

\section{Formulation of the Problem}\label{sec:2}
Consider the following Cornell potential

\begin{equation*}
V(r)=-\frac{A}{r}+B^{2}r
\end{equation*}
where $\mathit{A}$, $\mathit{B}$ are real and positive constants, and $r\in(0,\infty)$. If we
substitute $\mathit{V}(r)$ into Schr\"{o}dinger equation in three-dimensions, we have

\begin{equation}
\left\{ \frac{d^{2}}{dr^{2}}+\epsilon-\left[ -\frac{\alpha}{r}+\rho r+\frac{l(l+1)}{r^{2}%
}\right] \right\} \Psi (r)=0  \label{potsch}
\end{equation}
where $\epsilon=2\mu E_n$, $\alpha=2\mu A$ and $\rho=2\mu B^2$. $E_n$ and $\mu=\frac{m_{1}m_{2}}{m_{1}+m_{2}}$ are energy eigenvalue of \textit{n} th level and reduced mass of the q\={q} system, respectively ($m_{1}$ and $m_{2}$ are quark masses). After changing the variable, in Eq.(\ref{potsch}), as $r=u^{2}$, then substituting $\Psi
(u)=u^{1/2}g(u)$, we get\newline
\begin{equation}\label{afterr2u1/2}
g^{\prime\prime}(u)+\left[4\epsilon u^2+4\alpha-4\rho u^{4}- \frac{4l(l+1)+\frac{3}{4}}{u^{2}}  \right]g(u)=0 
\end{equation}

If one puts $g(z)=z^{\gamma+1}e^{-\frac{z^{3}}{3}}f(z)$ into Eq.(\ref{afterr2u1/2}), in accordance with the domain of the problem, we have  

\begin{equation}
f^{\prime \prime }(z)=2\left[ z^{2}-\frac{\gamma +1}{z}\right] f^{\prime
}(z)+[2(\gamma +2)z-\sigma z^{2}-\omega ]f(z)  \label{solab}
\end{equation}
where $\omega =\frac{4\alpha}{(4\rho)^{\frac{1}{3}}}$, $\sigma =\frac{4\epsilon}{(4\rho)^{\frac{2}{3}}}$, $\gamma=2l+\frac{1}{2}$ and $z=(4\rho)^{\frac{1}{6}}u$. The final equation is suitable for applying AIM. After this point, we can apply AIM to the
problem in two different ways: one is direct application (i.e., approximate solution) to get the numerical results and the other is usage of the method in scope of perturbation theory to obtain perturbative energies through a perturbation expansion as follows  
\begin{equation}
\sigma =\sigma _{0}+\omega \sigma _{1}+\omega ^{2}\sigma _{2}+...
\end{equation}
where $\sigma _{0}$, $\sigma _{1}$, $\sigma _{2}$,... are perturbation expansion coefficients. These can be obtained independently from the potential parameters. Thus, we can get a semi-analytical formula for the energy eigenvalues. One can also achieve the mass-energy of the system by using this formula, as given in the Section 4.


\subsection{Numerical Results}

In this section, we directly apply AIM to Eq.(\ref{solab}) to get the energy eigenvalues for different potential parameters, and we compare our results with the perturbative energies, for which Eq.(\ref{epertson}) in the next section has been used.

\begin{equation*}
f^{\prime \prime }(z)=2\left[ z^{2}-\frac{\gamma +1}{z}\right] f^{\prime
}(z)+[2(\gamma +2)z-\sigma z^{2}-\omega ]f(z)
\end{equation*}

From this equation, it is easily seen that $\lambda _{0}(z)=2\left[ z^{2}-\frac{\gamma +1}{z}%
\right] $ and $s_{0}(z)=2(\gamma +2)z-\sigma z^{2}-\omega $ according to Eq.(\ref{aimform}). We tabulate the results of direct application of AIM in
Table 1, Table 2 and Table 3. For simplicity, in the calculations, the reduced mass has been considered $\mu=\frac{1}{2}$. In Table 1 the potential parameters have been chosen as $A=B=1$ while $A=1$, $B=\frac{1}{10}$ in Table 2, and $A=1$, $B=10$ in Table 3. $E_{pert}$, seen in the tables, is for the comparison and has been obtained by using Eq.(\ref{epertson}).

\begin{table}[h]
	\caption{Comparisons of the perturbative energy eigenvalues with those obtained by direct application of AIM for the potential parameters $A=B=1$, and for the reduced mass $\mu=\frac{1}{2}$.}
	\centering
	\begin{tabular}{|c|c|c|c|c|c|}
		\hline
		$n$ & $E_{n0}$ & $E_{pert}$ & $l$ & $E_{0l}$ & $E_{pert}$ \\ \hline
		$0$ & $1.39788$ & 1.41015 & $0$ & $1.39788$ & 1.41015 \\ 
		$1$ & $3.47509$ & 3.47509 & $1$ & $2.82565$ & 2.8269 \\ 
		$2$ & $5.03291$ & 5.03224 & $2$ & $3.85058$ & 3.85089 \\ 
		$3$ & $6.37015$ & 6.36948 & $3$ & $4.72675$ & 4.72687 \\ 
		$4$ & $7.57493$ & - & $4$ & $5.51698$ & - \\
		$5$ & $8.68791$ & - & $5$ & $6.24840$ & - \\ \hline
	\end{tabular}
\end{table}

\begin{table}[h]
	\caption{Comparisons of the perturbative energy eigenvalues with those obtained by direct application of AIM for the potential parameters $A=1$, $B=\frac{1}{10}$, and for the reduced mass $\mu=\frac{1}{2}$.}
	\centering
	\begin{tabular}{|c|c|c|c|c|c|}
		\hline
		$n$ & $E_{n0}$ & $E_{pert}$ & $l$ & $E_{0l}$ & $E_{pert}$ \\ \hline
		$0$ & $-0.221031$ & -0.164433 & $0$ & $-0.221031$ & -0.164433 \\ 
		$1$ & $0.0347222$ & 0.033627 & $1$ & $0.0174006$ & 0.023501 \\ 
		$2$ & $0.141913$ & 0.138477 & $2$ & $0.102472$ & 0.104008 \\ 
		$3$ & $0.220287$ & 0.217229 & $3$ & $0.159831$ & 0.160406 \\
		$4$ & $0.286111$ & - & $4$ & $0.206238$ & - \\ 
		$5$ & $0.344602$ & - & $5$ & $0.246681$ & - \\ \hline
	\end{tabular}
\end{table}

\newpage

\begin{table}[h]
	\caption{Comparisons of the perturbative energy eigenvalues with those obtained by direct application of AIM for the potential parameters $A=1$, $B=10$, and for the reduced mass $\mu=\frac{1}{2}$.}
	\centering
	\begin{tabular}{|c|c|c|c|c|c|}
		\hline
		$n$ & $E_{n0}$ & $E_{pert}$ & $l$ & $E_{0l}$ & $E_{pert}$ \\ \hline
		$0$ & $46.4022$ & 46.4047 & $0$ & $46.4022$ & 46.4047 \\ 
		$1$ & $85.3393$ & 85.3394 & $1$ & $70.0161$ & 70.0165 \\ 
		$2$ & $116.729$ & 116.729 & $2$ & $89.7154$ & 89.7154 \\ 
		$3$ & $144.315$ & 144.315 & $3$ & $107.334$ & 107.334 \\ 
		$4$ & $169.461$ & - & $4$ & $123.562$ & - \\ 
		$5$ & $192.851$ & - & $5$ & $138.761$ & - \\ \hline
	\end{tabular}
\end{table}

\section{Perturbation Theory}\label{sec:4}

Although the usage of perturbation method in the frame of AIM is comprehensively introduced in \cite{pert},
we give a summary about the methodology in this section, assuming that the
potential of a system is written as

\begin{equation}
V(x)=V_{0}(x)+\mu V_{p}(x)  \label{pertpot}
\end{equation}
where $V_{0}(x)$ is solvable (unperturbed Hamiltonian) potential. $%
V_{p}(x)$ and $\mu $ are potential of the perturbed Hamiltonian and
perturbation expansion parameter, respectively. The Schr\"{o}dinger equation then reads,

\begin{equation}
\left( -\frac{d^{2}}{dx^{2}}+V_{0}(x)+\mu V_{p}(x)\right) \Psi (x)=E\Psi (x)
\label{pertsch}
\end{equation}
where $E_{n}$ eigenvalues are written as a series expansion of $j$ th-order
correction $E_{n}^{(j)}$ as follows:

\begin{equation}
E_{n}=E_{n}^{(0)}+\mu E_{n}^{(1)}+\mu
^{2}E_{n}^{(2)}+...=\sum\limits_{j=0}^{\infty }\mu ^{j}E_{n}^{(j)}
\label{perteig}
\end{equation}

After substituting $\psi (x)=\psi _{0}(x)f(x)$ in Eq.(\ref{pertsch}), one can obtain
the following equation for $f(x)$

\begin{equation}
f^{\prime \prime }(x)=\lambda _{0}(x,\mu ,E)f^{\prime }(x)+s_{0}(x,\mu
,E)f(x)  \label{pertAIM}
\end{equation}
and the termination condition in this case can be written as

\begin{equation}
\delta _{n}(x,\mu ,E)=s_{n}(x,\mu ,E)\lambda _{n-1}(x,\mu ,E)-\lambda
_{n}(x,\mu ,E)s_{n-1}(x,\mu ,E)=0  \label{pertdelta}
\end{equation}

Once $\delta _{n}(x,\mu ,E)$ is expanded about $\mu =0$, we obtain

\begin{equation}
\delta _{n}(x,\mu ,E)=\delta _{n}(x,0,E)+\left. \frac{\mu }{1!}\frac{%
	\partial \delta _{n}(x,\mu ,E)}{\partial \mu }\right\vert _{\mu =0}+\left. 
\frac{\mu ^{2}}{2!}\frac{\partial ^{2}\delta _{n}(x,\mu ,E)}{\partial \mu
	^{2}}\right\vert _{\mu =0}+...=\sum\limits_{k=0}^{\infty }\mu ^{k}\delta
_{n}^{(k)}(x,E)=0  \label{pertdeltaseri}
\end{equation}
where $\delta _{n}^{(k)}(x,E)=\left. \frac{1}{k!}\frac{\partial ^{k}\delta
	_{n}(x,\mu ,E)}{\partial \mu ^{k}}\right\vert _{\mu =0}$.

According to perturbation method in the framework of AIM, solving the
equation $\delta _{n}(x,0,E)=0$ with respect to (unknown) $E$ gives $%
E_{n}^{(0)}$ (eigenvalues of unperturbed Hamiltonian), equation $\delta
_{n}^{(1)}(x,E)=0$ gives $E_{n}^{(1)}$ (first-order correction to $E_{n}$), $%
\delta _{n}^{(2)}(x,E)$ gives $E_{n}^{(2)}$ (second-order correction to $%
E_{n}$) and so on. Besides, the perturbative eigenfunctions can be achieved in the same vein with the eigenvalues.
This is an alluring feature of the AIM usage in the perturbation theory
for obtaining the eigenfunctions $f_{n}(x)$ given as follows,

\begin{equation}
f_{n}(x)=\exp \left( -\int\limits^{x}\alpha _{n}(t,\mu )dt\right)
\label{perteigenfunc}
\end{equation}
where $\alpha _{n}(t,\mu )\equiv s_{n}(t,\mu )/\lambda _{n}(t,\mu )$. $%
\alpha _{n}(t,\mu )$ is expanded about $\mu =0$ in a similar manner, done for
obtaining the eigenvalues. So,

\begin{equation}
\alpha _{n}(t,\mu )=\sum\limits_{k=0}^{\infty }\mu ^{k}\alpha _{n}^{(k)}(t)
\label{pertalfa}
\end{equation}
where\ $\alpha _{n(x)}^{(k)}=\left. \frac{1}{k!}\frac{\partial ^{k}\alpha
	_{n(x,\mu )}}{\partial \mu ^{k}}\right\vert _{\mu =0}$. Thus, perturbation
expansion of the $f_{n(x)}$ is written as follows

\begin{equation}
f_{n}(x)=\exp \left[ \sum\limits_{k=0}^{\infty }\mu ^{k}\left(
-\int\limits^{x}\alpha _{n}^{(k)}(t)dt\right) \right] =\prod\limits_{k=0}^{%
	\infty }f_{n}^{(k)}(x)  \label{pertexpeigenfunction}
\end{equation}
where $k$ th-order correction $f_{n}^{(k)}(x)$ to $f_{n}(x)$ is

\begin{equation}
f_{n}^{(k)}(x)=\mu ^{k}\left( -\int\limits^{x}\alpha _{n}^{(k)}(t)dt\right)
\label{pertcorrectoeigenfunc}
\end{equation}

\subsection{Perturbation Theory for the Cornell Potential}

For our problem, we may apply the perturbation expansion which has been elucidated in previous section to the following differential equation

\begin{equation*}
f^{\prime \prime }(z)=2\left[ z^{2}-\frac{\gamma +1}{z}\right] f^{\prime
}(z)+[2(\gamma +2)z-\sigma z^{2}-\omega ]f(z)
\end{equation*}

Suppose that $\sigma$ is written as follows 

\begin{equation}
\sigma (n,l)=\sigma _{0}(n,l)+\sigma _{1}(n,l)\omega +\sigma _{2}(n,l)\omega
^{2}+...  \label{omegapert}
\end{equation}
where $\omega$ is the perturbation expansion parameter. So, the energy eigenvalue is yielded as

\begin{equation}\label{eunpert}
E_{pert}=\left( \frac{(4\rho)^{2/3}}{8\mu}\right) ^{\frac{2}{3}}\sigma (n,l)
\end{equation}
and more clearly

\begin{equation}\label{epert}
E_{pert}=\frac{(4\rho)^{2/3}}{8\mu}\sigma _{0}(n,l)+\frac{(4\rho)^{1/3}}{2\mu}\alpha\sigma
_{1}(n,l)+\frac{2\alpha^2}{\mu}\sigma _{2}(n,l)+...
\end{equation}

In the above expansion, the general form of the zeroth-order correction $\sigma _{0}$ is obtained via

\begin{equation}
\delta ^{(0)}(z,0,\sigma _{0})=0  \label{deltazerothorder}
\end{equation}

The first-order correction, $\sigma_1$, is obtained by using the equation $\delta ^{(1)}(z,0,\sigma _{1})=0$ in the same manner with the $\sigma_0$, while $\delta ^{(2)}(z,0,\sigma _{2})=0$ is used for $\sigma_2$. Numerical results of $\sigma_{0}$, $\sigma_{1}$ and $\sigma_{2}$ coefficients, obtained by AIM, are reported in Table 4 for some energy levels. Besides, for $\mu=\frac{1}{2}$, comparisons of the perturbative energy eigenvalues with the ones obtained by direct application of AIM have been given in Table 1, Table 2 and Table 3, in previous section. We emphasize, in Table 4, that corrections to the perturbation expansion do not depend on the potential parameters.

\begin{table}[h]
	\caption{Perturbation coefficients of the expansion given as Eq.(\ref{omegapert}) and Eq.(\ref{epert}). Notice that corrections to the perturbation expansion do not depend on the potential parameters.}
	\centering
	\begin{tabular}{|c|c|c|c|c|}
		\hline
		$l$ & $n$ & $\sigma _{0}(n,l)$ & $\sigma _{1}(n,l)$ & $\sigma _{2}(n,l)$ \\ 
		\hline
		\multirow{4}{0.5cm}{0} & $0$ & $3.71151$ & $-0.525933$ & $-0.0232729$ \\ 
		& $1$ & $6.48922$ & $-0.366743$ & $-0.00767365$ \\ 
		& $2$ & $8.76334$ & $-0.297538$ & $-0.00400191$ \\ 
		& $3$ & $10.7732$ & $-0.256486$ & $-0.00251618$ \\ \hline
		\multirow{4}{0.5cm}{1} & $0$ & $5.33566$ & $-0.322683$ & $-0.00554189$ \\ 
		& $1$ & $7.75358$ & $-0.258925$ & $-0.00282569$ \\ 
		& $2$ & $9.85399$ & $-0.222298$ & $-0.00176295$ \\ 
		& $3$ & $11.7558$ & $-0.197751$ & $-0.00122526$ \\ \hline
		\multirow{4}{0.5cm}{2} & $0$ & $6.74357$ & $-0.244191$ & $-0.00241586$ \\ 
		& $1$ & $8.93661$ & $-0.208300$ & $-0.00148846$ \\ 
		& $2$ & $10.9037$ & $-0.184664$ & $-0.00102765$ \\ 
		& $3$ & $12.7146$ & $-0.167585$ & $-0.000761053$ \\ \hline
		\multirow{4}{0.5cm}{3} & $0$ & $8.01784$ & $-0.200753$ & $-0.00134507$ \\ 
		& $1$ & $10.0516$ & $-0.177251$ & $-0.000921458$ \\ 
		& $2$ & $11.9129$ & $-0.160449$ & $-0.000679139$ \\ 
		& $3$ & $13.6471$ & $-0.147666$ & $-0.000525832$ \\ \hline
	\end{tabular}
\end{table}

As can be seen from Tables 1-3, the perturbative energy eigenvalues are in very good agreement with the numerically obtained ones, even for small values of the parameter $B$. Furthermore, they are in accordance to each other for $B\geq 1$, while $A=1$ (see in Table 1 and Table 3). Additionally, this agreement is much better for higher quantum states. The perturbative eigenvalues are a little bit different from that obtained as numerically, for $B<1$, $A=1$ and the lower quantum states (see in Table 2). However, they are in agreement for the higher levels.

\newpage

As a practice, we have applied our perturbation expansion formula (up to second-order correction) to get the ground-state energies of quarkonium in Table 5, for various values of the parameter $A$, while $B=1$ and $\mu=\frac{1}{2}$. In Table 5, we also report comparisons of the perturbative energy eigenvalues with the ones of s-wave heavy quarkonium from Refs.\cite{chung} and \cite{hallnsaad}

\begin{table}[h]
	\caption{Comparisons of energy eigenvalues obtained by using the perturbation expansion formula in Eq.(\ref{epert}) ($E_{pert}$) with the ones of s-wave heavy quarkonium from Refs.\cite{chung} and \cite{hallnsaad}. The potential paramater $B$ is taken as $B=1$, while the reduced mass is $\mu=\frac{1}{2}$ in this case. The eigenvalues of Refs.\cite{chung} and \cite{hallnsaad} are exact results.}
	\centering
	\begin{tabular}{|c|c|c|c|c|c|c|c|}
		\hline
		$A$ & $E_{00}$(Ref.\cite{chung}) & $E_{00}$(Ref.\cite{hallnsaad}) & $E_{pert}$ & $A$ & $E
		_{00}$(Ref.\cite{chung}) & $E_{00}$(Ref.\cite{hallnsaad}) & $E_{pert}$ \\ \hline
		0.2 & 2.16732 & 2.16732 & 2.16741 & 0.1 & 2.25368 & 2.25368 & 2.25369 \\ 
		\hline
		0.4 & 1.98850 & 1.98850 & 1.98923 & 0.3 & 2.07895 & 2.07895 & 2.07927 \\ 
		\hline
		0.6 & 1.80107 & 1.80107 & 1.80367 & 0.5 & 1.89590 & 1.89590 & 1.89740 \\ 
		\hline
		0.8 & 1.60441 & 1.60441 & 1.61063 & 0.7 & 1.70394 & 1.70393 & 1.70808 \\ 
		\hline
		1 & 1.39788 & 1.39788 & 1.41015 & 0.9 & 1.50242 & 1.50242 & 1.51132 \\ \hline
		1.2 & 1.18084 & 1.18083 & 1.20221 & 1.1 & 1.29071 & 1.29071 & 1.30711 \\ 
		\hline
		1.4 & 0.95264 & 0.95264 & 0.98683 & 1.3 & 1.06817 & 1.06817 & 1.09545 \\ 
		\hline
		1.6 & 0.71266 & 0.71266 & 0.76400 & 1.5 & 0.83416 & 0.83416 & 0.87635 \\ 
		\hline
		1.8 & 0.46027 & 0.46026 & 0.53373 & 1.7 & 0.58805 & 0.58805 & 0.64980 \\ 
		\hline
	\end{tabular}
\end{table}

\newpage

As is seen from Table 5, the results for which our perturbation expansion (up to second-order correction) has been used are in very good agreement with Refs.\cite{chung} and \cite{hallnsaad} for small values of $A$. However, our analytical results are little bit different from the exact ones as the $A$ gets larger values. It seems that the perturbation expansion, which includes third-order correction, may give more accurate results. The more correction term we add to the perturbative expansion, the more compatible results we get. Nevertheless, we can say that Eq.(\ref{epert}) can be used as an eigenvalue formula of the Schr\"{o}dinger equation in case of Cornell potential, for practical purposes. So, one can use the following formula

\begin{equation}
E_{pert}=\frac{(4\rho)^{2/3}}{8\mu}\sigma _{0}(n,l)+\frac{(4\rho)^{1/3}}{2\mu}\alpha\sigma
_{1}(n,l)+\frac{2\alpha^2}{\mu}\sigma _{2}(n,l) \label{epertson}
\end{equation}
for obtaining the eigenvalues and mass of the quarkonium for Cornell potential. Besides, it can be fit to mass formula of experimental values for determining the potential parameters $A$ and $B$. The advantage of Eq.(\ref{epertson}) is that the coefficients $\sigma _{0}$, $\sigma _{1}$ and $\sigma _{2}$ are independent of the potential parameters.

\subsection{Energy Eigenvalues and Mass Spectrum for Heavy Quarkoniums}

In this section, we tested our formula through cross-checking with the exact results in literature and with the experimental data. For comparing our energy eigenvalues with the exact ones, the parameters of Cornell potential have been considered $A=0.52$ and $B=0.43$. Besides, we have chosen the quark masses as $m_c=1.84$ GeV and $m_b=5.18$, in this case \cite{alfredo}.

Also, we tested our formula by comparing our results, for the masses of heavy mesons, with the experimental data. For doing this, we have taken the quark masses as $m_c=1.44$ GeV and $m_b=4.87$ GeV, and the potential parameters as $A=0.64$ and $B=0.39$. All these values have been obtained by fitting our formula to the experimental data in Ref.\cite{pdg}.  

In Table 6, we compared our energy eigenvalues calculated by using Eq.(\ref{epertson}) with the ones of Ref.\cite{alfredo}. Furthermore, in Table 7, we gave our results for the masses of the mesons obtained by the same equation. Table 7 also includes the experimental data got from Ref.\cite{pdg}.

\begin{table}[h]
	\caption{Comparisons of the energy eigenvalues (in GeV) of the mesons c\={c}, b\={b} and b\={c} calculated by using Eq.(\ref{epertson}) with the exact ones of Ref.\cite{alfredo}. The parameters of Cornell potential are $A=0.52$ and $B=0.43$, while the quark masses are $m_c$=1.84 GeV and $m_b$=5.18 GeV.} 
	\centering
	\begin{tabular}{cccccccccc}
		\hline
		 \multirow{2}{0.7cm}{}& \multicolumn{3}{c}{c\={c}} & \multicolumn{3}{c}{b\={b}} & \multicolumn{3}{c}{b\={c}} \\
		 \hline
		 $E_n$ & Exact \cite{alfredo} & Ref.\cite{alfredo} & AIM & Exact \cite{alfredo} & Ref.\cite{alfredo} & AIM & Exact \cite{alfredo}& Ref.\cite{alfredo} & AIM \\
		 \hline
		 1s& 0.2575 & 0.2578 & 0.2660 & -0.1704 & -0.1702 & -0.1216 & 0.1110 & 0.1113 & 0.1269 \\
		 2s & 0.8482 & 0.8096 & 0.8481 & 0.4214 & 0.3579 & 0.4203 & 0.6813 & 0.6324 & 0.6803 \\
		 3s & 1.2720 & 1.1427 & 1.2715 & 0.7665 & 0.5612 & 0.7635 & 1.0686 & 0.9065 & 1.0668 \\
		 \hline	 
	\end{tabular}
\end{table}

\begin{table}[h]
	\caption{Comparisons of the masses (in GeV), obtained via AIM, of the heavy mesons c\={c}, b\={b} and c\={b} with the ones of Ref.\cite{alfredo}, and with the experimental data from \cite{pdg}. In this case, we have taken the quark masses as $m_c$=1.44 GeV and $m_b$=4.87 GeV, and the potential parameters as $A=0.64$ and $B=0.39$, for our calculations. All these parameters have been obtained by fitting our formula, given in Eq.(\ref{epertson}), to the experimental data.}
	\centering
	\begin{tabular}{ccccccccccccc}
		\hline
		\multirow{2}{0.7cm}{}& \multicolumn{3}{c}{c\={c}} & \multicolumn{3}{c}{b\={b}} & \multicolumn{3}{c}{c\={b}} \\
		\hline
		$M_n$ & Exp.& Ref.\cite{alfredo} & AIM & Exp. & Ref.\cite{alfredo} & AIM &Exp.& Ref.\cite{alfredo} & AIM \\
		\hline
		1s & 3.097 & 3.097 & 3.096 & 9.460 &9.350&9.462&6.275 &6.291&6.362 \\
		2s & 3.686& 3.649 &3.672 & 10.023 & 9.878 & 10.027 & 6.842 & 6.812 & 6.911 \\
		3s & 4.039 & 3.963 & 4.085 & 10.355 & 10.081 & 10.361 & - & 7.087 & 7.284 \\
		4s&-&-&4.433&10.579&-&10.624&-&-&7.593\\
		1p&3.511&-&3.521&9.899&-&9.963&-&-&6.792\\
		2p&3.927&-&3.951&10.260&-&10.299&-&-&7.178\\3p&-&-&4.310&10.512&-&10.564&-&-&7.494\\
		1d&-&-&3.800&10.164&-&10.209&-&-&7.051\\ 
		\hline			 
	\end{tabular}
\end{table}

It can be seen from Table 6 that the energy eigenvalues of the mesons c\={c}, b\={b} and b\={c}, obtained by Eq.\ref{epertson}, are more compatible with the exact ones, than those of Ref.\cite{alfredo}. The difference between AIM and Ref.\cite{alfredo} becomes clearer as the energy level increases. Similar things can be said for the masses in Table 7: the results obtained via AIM are closer to the experimental data than those of Ref.\cite{alfredo}.

\section{Conclusion}\label{sec:5}

We have used AIM to obtain both, the eigenvalues of Schr\"{o}dinger equation and mass of q\={q} system for Cornell potential, in three-dimensions. AIM has some advantages such as being used for either exactly or numerically (or approximately) solvable problems. Furthermore, one can use AIM in the frame of perturbation theory. Once it is performed to obtain perturbative solutions, the wavefunction of unperturbed Hamiltonian is not needed to get the corrections to the perturbation expansion.

In the present study, the energy eigenvalues in the case of Cornell potential have been achieved by direct application of the method. Besides, we have performed perturbation theory in the view of AIM for the problem and found a semi-analytical formula for energy eigenvalues. Numerical results obtained by using this formula, for the reduced mass $\mu=\frac{1}{2}$, conform with the exact results of Refs.\cite{chung,hallnsaad}, in a wide spectrum of the potential parameters $A$ and $B$ (especially for $B>A$). Furthermore, the results are compatible with the ones obtained directly, in Section 3. It is also possible to see from the results that the perturbative eigenvalues fit in with the exact ones for higher quantum states, even for the large values of $A$. For any values of $A$ and $B$, the higher quantum states are more consonant with the exact ones than the lower states. The perturbation expansion, which includes third-order correction, may give more accurate results. The more correction terms we add to the perturbative expansion, the more compatible results we may get.

We have also tested our semi-analytical formula, by cross-checking it with the exact results in literature, and with the experimental data. It can be seen, from Table 6, that our energy eigenvalues calculated by using Eq.(\ref{epertson}) are more compatible with the exact ones than those of Ref.\cite{alfredo}. Furthermore, the difference between our results and Ref.\cite{alfredo} becomes clearer as the energy level increases. By using AIM, we have also obtained mass results which are closer to the experimental data than Ref.\cite{alfredo}.

As a consequence, semi-analytical formula achieved for energy eigenvalues and mass of quarkonium can be used for practical purposes in the case of Cornell potential. If our formula is fitted to the experimental data, the potential parameters (and masses of the quarks, if it is needed) can also be obtained.

\bigskip

\noindent\textbf{Acknowledgement}

This academic work was supported by the Mersin Technology Transfer Office Academic Writing Center of Mersin University.

\end{document}